\documentclass[12pt]{article}
\usepackage{graphicx}  
\usepackage{dcolumn}   
\usepackage{bm}        
\usepackage{amssymb}   
\usepackage{color}

\begin{document}
\pagestyle{plain}
\title{\bf A spatial analog of the Ruelle-Takens-Newhouse scenario 
developing in reactive miscible fluids}
\author{Dmitri Bratsun\\
Department of Applied Physics,\\
Perm National Research Polytechnical University,\\
Perm, 614990, Russia\\
E-mail: dmitribratsun@rambler.ru}

\maketitle
\thispagestyle{empty}

\pagestyle{plain}
\begin{abstract}
{\noindent
We present a theoretical study on pattern formation occurring in miscible 
fluids reacting by a second-order reaction $A + B \to S$ in a vertical Hele-Shaw 
cell under constant gravity. We have recently reported that concentration-dependent 
diffusion of species coupled with a frontal neutralization reaction can produce 
a multi-layer system where low density depleted zones could be embedded 
between the denser layers. This leads to the excitation of chemoconvective modes 
spatially separated from each other by a motionless fluid. In this paper, 
we show that the layers can interact via a diffusion mechanism. 
Since diffusively-coupled instabilities initially have different wavelengths, this causes 
a long-wave modulation of one pattern by another. We have developed 
a mathematical model which includes a system of reaction-diffusion-convection equations. 
The linear stability of a transient base state 
is studied by calculating the growth rate of the Lyapunov exponent for each unstable layer. 
Numerical simulations supported by the phase portrait reconstruction and Fourier 
spectra calculation have revealed that nonlinear dynamics consistently passes 
through (i) a perfect spatially periodic system of chemoconvective cells; 
(ii) a quasi-periodic system of the same cells; (iii) a disordered fingering structure. 
We show that in this system, the coordinate co-directed to the reaction front 
paradoxically plays the role of time, time itself acts as a bifurcation parameter, 
and a complete spatial analog of the Ruelle-Takens-Newhouse scenario 
of the chaos onset is observed.}
\end{abstract}

In recent decades, the study of the interaction between reaction-diffusion 
phenomena and convective instabilities brought many surprises. 
Let us focus on a neutralization reaction: a second-order $A + B \to C$ 
reaction is distinguished by a comparatively simple, albeit nonlinear, kinetics. 
If two species are initially separated in space, the reaction proceeds 
in a frontal manner due to the high value of the reaction rate constant.
In this case, it may result in various buoyancy-driven instabilities 
if the reaction occurs either in immiscible two-layer 
systems~\cite{eckert99,bra04,eckert04,eckert08,bra11} 
or in miscible acid-base systems~\cite{ron08,zal08,alma10,mul12,hej12,carb13,kim14,bra15,trev15,bra16,bra17}. 
Among the important effects that have been observed here, we highlight 
the pattern convection in the form of a perfectly periodic system of cellular-like 
fingers keeping contact with the interface when an organic solvent containing 
an acid $A$ is in contact with an aqueous solution of an inorganic base 
B~\cite{eckert04}. At that time, a liquid-liquid interface has been recognized 
as the main reason for this unusual regularity of salt fingering~\cite{bra04,bra11}.
However, then a perfectly organized structure of fingers 
has been observed already in the miscible system, where two aqueous solutions 
of base and acid have been brought into contact~\cite{bra15,bra16}.
It was shown that this pattern arises due to the strong dependence 
of the diffusion coefficients of the initial reactants and the reaction product 
on their concentrations. The effect of concentration-dependent 
diffusion (hereinafter CDD) coupled with a fast neutralization reaction
has been demonstrated to produce a spatially localized zone with unstable 
density stratification (figuratively, a density pocket) in a system with inherently 
stable configuration, when a less dense solution is placed above a more dense one. 
It should be noted that such an effect creates a new situation in fluid mechanics
when the convective modes arise in different parts of the medium and compete 
on the distance. In this case, the patterns can be coupled by the diffusion 
of heat or matter which transmit a signal through an interlayer 
of the motionless fluid.

Generally, the CDD effect means a significant expansion of the degrees of freedom
for the system to produce various types of instabilities and nonlinear 
dynamics that do not fit into the traditional classification~\cite{trev15}. 
For example, we have shown recently that when varying initial concentrations 
of solutions, the density pocket may collapse suddenly, causing a density shock wave 
separating the fluid at rest and the area with anomalously 
intense convective mixing~\cite{bra17}.

In this communication, we study the nonlinear interaction between 
two periodic systems of chemoconvective cells that arise independently
inside two different layers low in density. 
Although initially the layers are separated by the motionless fluid, 
they can nevertheless influence each other via the diffusion of the reactants. 
Thus, this configuration reproduces 
the conditions for the realization of a spatial analog 
of the Ruelle-Takens-Newhouse scenario of the transition to chaos. 
As it is known, the Ruelle-Takens-Newhouse theory~\cite{ruel71,new78} 
postulates that the power spectrum of a dynamical system 
evolves as a function of the control parameter and at the beginning consists 
of one frequency, then two (sometimes three), and then the broadband 
noise characteristic of chaos should start to appear.
In our case, we demonstrate that a similar sequence of bifurcations 
is reproduced, but the role of time is played by the spatial horizontal coordinate 
with time itself now being a control parameter.

{\it Mathematical model}. -- We consider two aqueous solutions of acid $A$
and base $B$ filling the cavity, which is strongly compressed in one 
of directions, so that the Hele-Shaw (HS) approximation is applied.
Right after the process starts, the reagents with initial concentrations 
$A_0, B_0$ diffuse into each other and are neutralized at the rate $k$ 
with the formation of salt $C$. Since both reagents are dissolved in water, 
their mixing begins at the contact of initially separated layers.
It is noteworthy that the problem is non-autonomous because reagents 
are not replenished during the reaction and concentration profiles change irreversibly.
Generally, a neutralization reaction is known to be exothermic.
However, in this paper, the heat release is neglected. 
In experiments, one can always achieve such reaction conditions, 
if the walls of the HS cell are made quite conductive for heat to dissipate.
In what follows, we assume that $A_0=B_0$. 

The system geometry is given by a two-dimension domain defined 
by $0\leqslant x\leqslant L$, $-H\leqslant z\leqslant H$ with $x$-axis 
directed horizontally and $z$-axis anti-directed to gravity. 
We scale the problem by using $2d$, $4d^2/D_{a0}$,
$D_{a0}/2d$, $A_0$ as the length, time, velocity and concentration 
scales, respectively. $D_{a0}$, $\nu$, $c_p$, $d$
stands for acid diffusivity (table value), kinematic viscosity 
and the HS semi-gapwidth.

\begin{figure}
\includegraphics [scale=0.34] {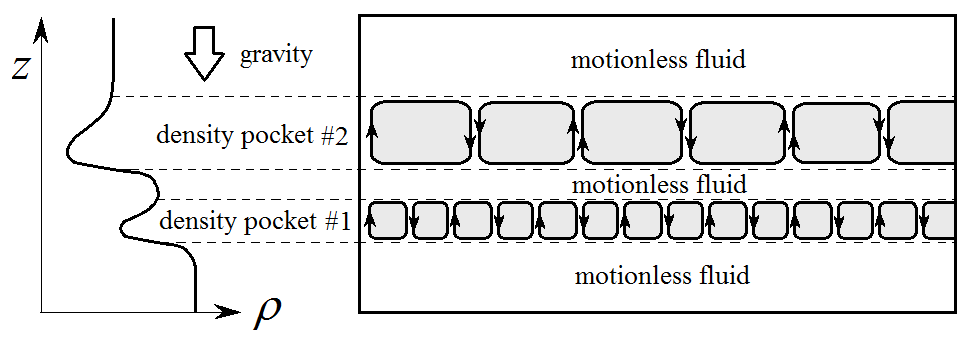}
\caption{\label{fig:1} Schematic presentation of two system 
of convective rolls independently arising in a depleted zone 
low in density.}
\end{figure}

The mathematical model
we develop consists in the set of equations
for species coupled to Navier-Stokes equation
written in the HS approximation~\cite{bra15}:
\begin{eqnarray}
\Phi + \nabla^2\Psi = 0,
\label{eq:1}
\\
\frac{1}{Sc}\left(\partial_t \Phi+\frac{6}{5}J(\Psi,\Phi)\right) = 
\nabla^2\Phi - 12\Phi  - \partial_x \rho ,
\label{eq:2}
\\
\partial_t A + J(\Psi,A) = \nabla (D_a (A) \nabla A) - \alpha AB ,
\label{eq:3}
\\
\partial_t B + J(\Psi,B) = \nabla (D_b (B) \nabla B) - \alpha AB ,
\label{eq:4}
\\
\partial_t C + J(\Psi,C) = \nabla (D_c (C) \nabla C) + \alpha AB ,
\label{eq:5}
\\
\rho = R_a A + R_b B + R_c C ,
\label{eq:6}
\end{eqnarray}
where $J(F,P)\equiv \partial_z F\partial_x P - \partial_x F\partial_z P$
stands for the Jacobian determinant. Here we use a two-field formulation 
for the motion equation, and introduce the stream function $\Psi$ 
and vorticity $\Phi$. The diffusion terms in Eqs.~(\ref{eq:3}-\ref{eq:5}) 
are written in the most general form, allowing the concentration-dependent
phenomena~\cite{crank75}. The problem involves following dimensionless 
parameters:
\begin{eqnarray}
Sc=\frac{\nu}{D_{a0}},\quad\alpha =\frac{4 k A_0 d^2}{D_{a0}},
\quad R_i = \frac{8 g\beta_i A_0 d^3}{\nu D_{a0}},
\label{eq:7}
\end{eqnarray}
which are the Schmidt number, the Damk\"ohler number, and
the set of solutal Rayleigh numbers, where $i=\{a,b,c\}$, respectively.
Eqs.~(\ref{eq:1}-\ref{eq:6}) should be supplemented 
by the boundary conditions
\begin{eqnarray}
x=0,L:\;\Psi=\partial_x\Psi =0,\;\partial_x A =0,\;\partial_x B =0,\;\partial_x C =0
\nonumber
\\
z=\pm H:\;\Psi=\partial_z\Psi =0,\;\partial_z A =0,\;\partial_z B =0,\;\partial_z C =0 
\label{eq:8}
\end{eqnarray}
and the initial conditions at $t=0$: 
\begin{eqnarray}
z\leqslant 0:\qquad \Psi=0,\quad A=0,\quad B=1; 
\nonumber
\\
z>0:\qquad \Psi=0,\quad A=1,\quad B=0.
\label{eq:9}
\end{eqnarray}

We have shown in~\cite{bra15,bra16} that the coupling between a second 
order reaction and nonlinear diffusion can transform an initially stably 
stratified fluid layer to a multi-layered system where the depleted 
zones low in density are embedded between the denser layers. 
This situation is presented schematically in Fig.~\ref{fig:1}. 
The figure shows a vertical density profile with two local minima, 
which can be defined as low-density pockets. 
It is important to note that the fluid layer shown in Fig.~\ref{fig:1} 
remains globally stable, since locally unstable fluid sublayers are not able 
to set in motion the adjacent immobile fluid. By changing the type of chemical reaction, 
the involved reagents, or their initial concentrations, we can 
create quite diverse configurations of the vertical stratification of the system.
In order to proceed further, it is necessary to specify closed-form exact 
laws of diffusion in~(\ref{eq:3}-\ref{eq:5}). The problem is that so 
far the CDD effect has been underestimated in fluid mechanics. Therefore, 
data on the concentration-dependence of the diffusion coefficients 
has appeared to be fragmentary and incomplete for most substances.
In what follows we use the diffusion laws developed recently 
for the pair HNO$_3$/ NaOH (for more details, see~\cite{bra15}): 
\begin{eqnarray}
D_a(A) \approx 0.881 + 0.158 A ,\nonumber
\\
D_b(B) \approx 0.594 - 0.087 B ,
\label{eq:10}
\\
D_c(C) \approx 0.478 - 0.284 C .\nonumber
\end{eqnarray}
The values of the parameters~(\ref{eq:7}) for the pair HNO$_3$/ NaOH 
can be estimated as follows: $Sc=317$, $\alpha=10^3$, $R_a = 3.2\times 10^5$, 
$R_b = 3.8\times 10^5$, $R_c = 5.1\times 10^5$. 

\begin{figure}
\includegraphics [scale=0.65] {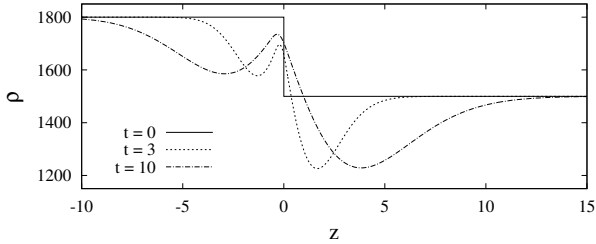}
\caption{\label{fig:2} 
Instantaneous base state profiles of the density $\rho$ 
are plotted against the vertical axis $z$ at times t = 0, 3, 10.}
\end{figure}

{\it Dynamics of the base state and its stability}. -- The system 
of equations~(\ref{eq:1}-\ref{eq:10}) allows the base state 
that describes the dynamics of pure reaction-diffusion processes. 
In this state, the fluid is at rest all the time.
We assume that the fluid velocity is zero and concentration fields 
depends only on the vertical coordinate $z$ and time $t$: 
$A_0(t,z)$, $B_0(t,z)$, $C_0(t,z)$. Then we obtain:
\begin{eqnarray}
\partial_t A_0 = D_a (A_0)\partial_{zz} A_0 + \partial_z D_a (A_0)
\partial_{z} A_0 - \alpha A_0 B_0 ,
\nonumber
\\
\partial_t B_0 = D_b (B_0)\partial_{zz} B_0 + \partial_z D_b (B_0)
\partial_{z} B_0 - \alpha A_0 B_0 ,
\label{eq:11}
\\
\partial_t C_0 = D_c (C_0)\partial_{zz} C_0 + \partial_z D_c (C_0)
\partial_{z} C_0 + \alpha A_0 B_0 .
\nonumber
\end{eqnarray}
The problem~(\ref{eq:8}-\ref{eq:11}) can be solved 
only numerically. 

Fig.~\ref{fig:2} shows the base state profiles of the density $\rho (t,z)$ 
defined by~(\ref{eq:6}) for three consecutive times $t=0,3,10$.
The system starts to evolve from the initial state, in which 
the lighter acid solution ($z\geqslant 0$) is above 
the more dense base solution ($z<0$). 
Thus, there exists a stable vertical stratification in terms of density, 
which excludes the development of Rayleigh-Taylor instability.
As soon as reaction-diffusion processes begin, the density profile undergoes 
dramatic changes: now it has two minima located above and below 
the reaction front implying a possible development of local instabilities 
in the depleted zones low in density. Thus, the formal scheme shown in Fig.~\ref{fig:1} 
is reproduced in practice. The mechanism of the formation 
of such a multi-layered system has been discussed in detail in~\cite{bra15,bra16}. 
Here we just briefly note that the main reason is that the reaction product
starts to be deposited near the reaction front. Since the diffusion 
coefficient of salt decreases with the growth of its concentration (the CDD effect), 
it can progressively accumulate near the reaction front making a local 
maximum in the density profile (Fig.~\ref{fig:2}, $t=3$). 
Since the acid has a higher value of the diffusion coefficient, over time, 
this maximum slowly shifts down (Fig.~\ref{fig:2}, $t=10$).
Thus, the diffusion makes the system to be non-autonomous, 
and time is the control parameter of the system.

\begin{figure}
(a)\includegraphics[scale=0.65]{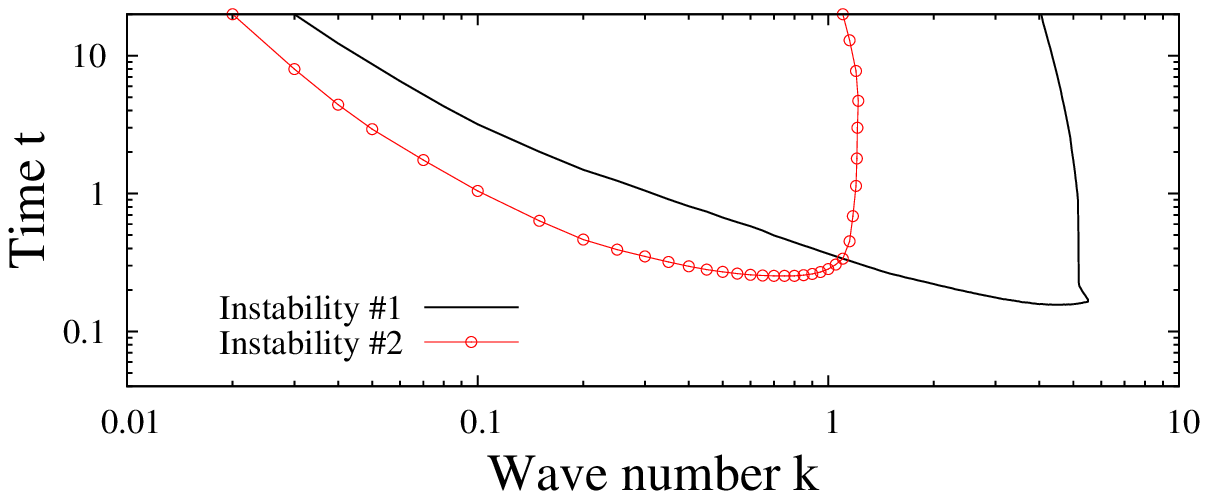}
(b)\includegraphics[scale=0.65]{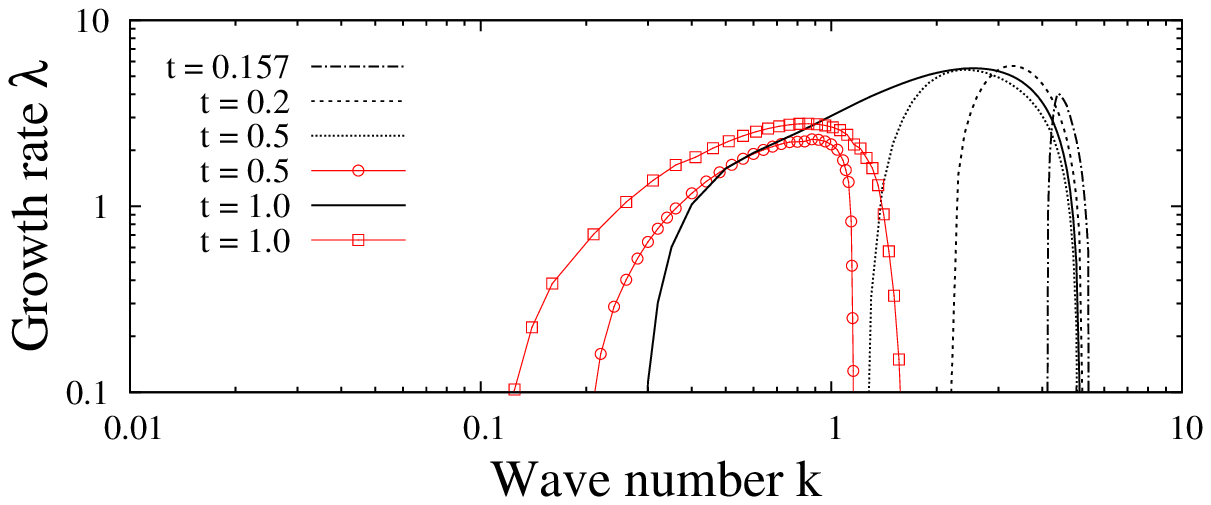}
\caption{\label{fig:3} (color online). (a) Neutral curves for two instabilities 
which arise in two zones low in density shown in Fig.~\ref{fig:2};
(b)~Real part of the growth rate of the instability $\#1$ (lines) 
and $\#2$ (line-points) are illustrated at different times.}
\end{figure}

Let us analyze the stability of a time-dependent base state 
defined by Eqs.~(\ref{eq:10},\ref{eq:11}) 
with respect to small monotonic perturbations:
\begin{equation}
\left(
\begin{array}{c}
\Phi(t,x,z)\\
\Psi(t,x,z)\\
A (t,x,z)\\
B (t,x,z)\\
S (t,x,z)
\end{array}\right) = 
\left(
\begin{array}{c}
0\\
0\\
A_0 (t,z)\\
B_0 (t,z)\\
S_0 (t,z)
\end{array}\right) +
\left(
\begin{array}{c}
\varphi (t,z)\\
\psi (t,z)\\
a (t,z)\\
b (t,z)\\
c (t,z)
\end{array}\right) e^{Ikx},
\label{eq:12}
\end{equation}
where $\varphi,\psi, a, b, c$ are, respectively, the amplitudes of normal 
form perturbations for the vorticity, stream function, acid, base and salt concentrations 
while $k$ is their wavenumber.
Substituting~(\ref{eq:12}) into Eqs.~(\ref{eq:1}-\ref{eq:6}) and linearizing 
these equations near the base state~(\ref{eq:11}), we obtain 
the following system of time-dependent amplitude equations for
the determination of critical perturbations:
\begin{eqnarray}
\phi + \partial_{zz}\psi - k^2\psi =0 ,
\nonumber
\\
\frac{1}{Sc}\partial_t \phi = \partial_{zz}\phi - (k^2+12)\phi 
- k^2 (R_a a + R_b b + R_c c) ,
\nonumber
\\
\partial_t a = D_a(A_0)(\partial_{zz}a - k^2 a) + D_a^{'} (2\partial_z A_0\partial_z a +
a\partial_{zz} A_0) 
\nonumber
\\
- \alpha (A_0 b + a B_0) -\psi\partial_z A_0,
\nonumber
\\
\partial_t b = D_b(B_0)(\partial_{zz}b - k^2 b) + D_b^{'} (2\partial_z B_0\partial_z b +
b\partial_{zz} B_0) 
\nonumber
\\
- \alpha (A_0 b + a B_0) -\psi\partial_z B_0,\;\;\;
\label{eq:13}
\\
\partial_t c = D_c(C_0)(\partial_{zz}c - k^2 c) + D_c^{'} (2\partial_z C_0\partial_z c +
c\partial_{zz} C_0) 
\nonumber
\\
+ \alpha (A_0 b + a B_0) -\psi\partial_z C_0,
\nonumber
\end{eqnarray}
where the linearization is carried out taking into account the explicit form 
of the diffusion laws~(\ref{eq:10}).

In order to determine the stability of a time-dependent base state,
we apply the method of the initial value problem (IVP). 
As it was shown in~\cite{homsy}, the IVP calculation
gives essentially the same results as the quasi-steady-state
approach, except for a short period of time in which the base
state changes rapidly. In our case, the system becomes unstable 
only after some critical period of time and the IVP usage is reasonable~\cite{bra04}.  
Thus, Eqs.~(\ref{eq:13}) are numerically integrated together 
with the base state problem~(\ref{eq:8}-\ref{eq:11}) and 
the boundary conditions for disturbances
\begin{eqnarray}
z=\pm H:\quad\phi=0,\;\partial_z a =0,\;\partial_z b =0,\;\partial_z c =0 
\label{eq:14}
\end{eqnarray}
to compute the growth rate $\lambda$ defined similarly to the Lyapunov exponent:
\begin{equation}
\lambda (t) = \frac{1}{N}\sum^N_{j=1}\frac{1}{\Delta t} \ln\frac{a_{j}
(t+\Delta t, z_{min})}{a_{j}(t, z_{min})},
\label{eq:15}
\end{equation}
where $\Delta t$ is the integration time step and $N$ is 
the number of independent realizations (typically 10-15). 
Because the growth rate $\lambda$ is sensitive to the given initial 
data, each independent integration started from white noise with an amplitude less 
than $10^{-4}$. We have fixed the occurrence of instability to the time when 
$\lambda (t)$ averaged over $N$ realizations changes sign 
from negative to positive. The position $z_{min}$ for measuring $\lambda$ 
was chosen at one of the local minima shown in Fig.~\ref{fig:2}.

\begin{figure}
\includegraphics[scale=0.18]{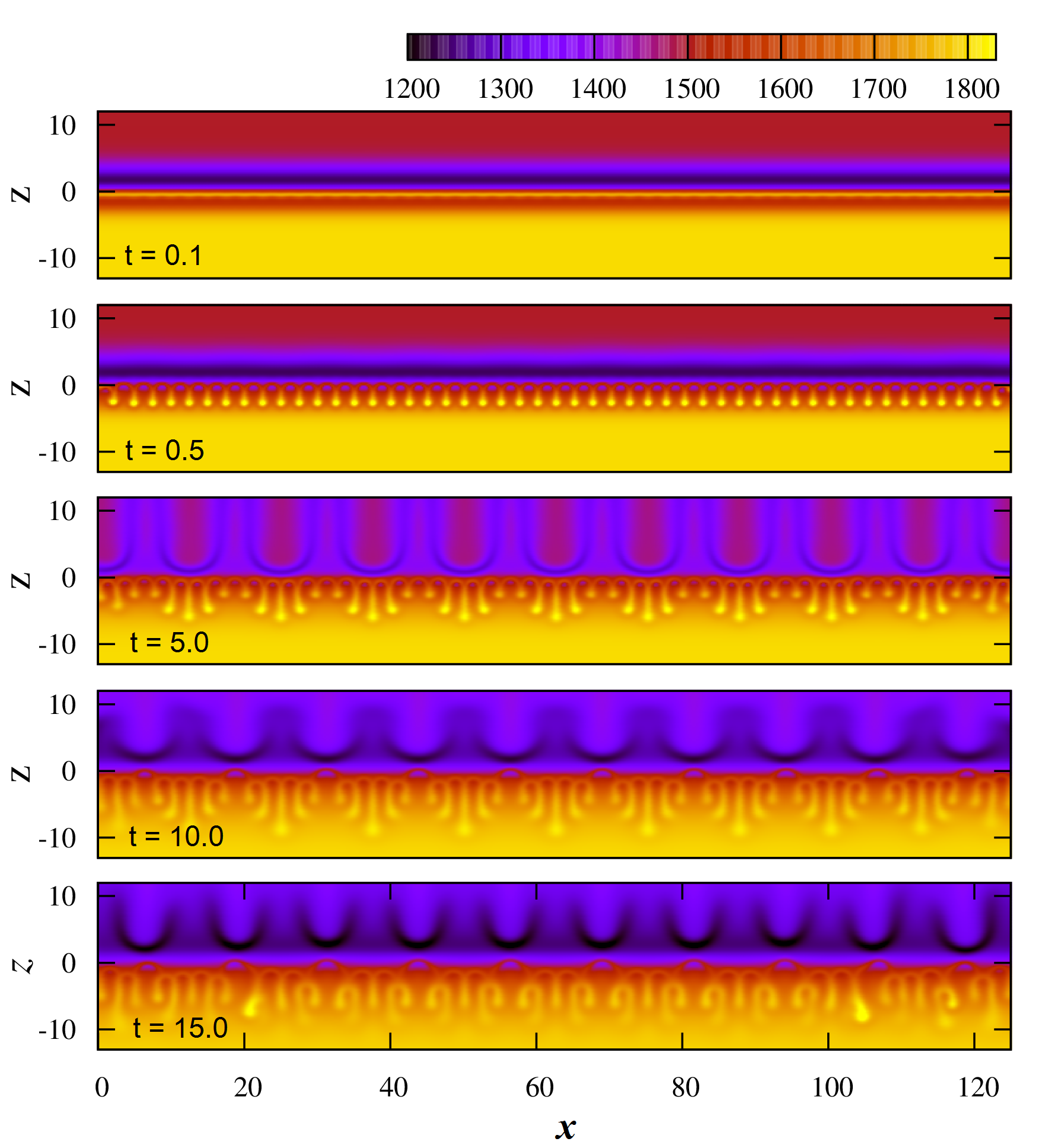}
\caption{\label{fig:4} (color online). 
Nonlinear evolution of the total density $\rho$ with time.  
The frames from up to down pertain to times $t = 0.1$, $0.5$, $5$, $10$, $15$ 
respectively. The domain of integration is $0\leqslant x\leqslant 125$, 
$-20\leqslant z\leqslant 20$.}
\end{figure}

Fig.~\ref{fig:3}a presents neutral curves for two instabilities which arise 
independently. The instability $\#1$ below the reaction front 
(in fact, the CDD instability) develops first. The minimum of the neutral 
curve corresponds to the wave number $k_1\approx 4.4$ at time 
$t_1\approx 0.156$. The instability $\#2$ 
starts at $k_2\approx 0.75$ and $t_2\approx 0.253$.
Fig.~\ref{fig:3}b presents the instantaneous growth rates $\lambda$ 
as a function of $k$ calculated for both instabilities.
Although the sublayers are separated by the immobile fluid, 
the instabilities still can influence each other through 
a diffusion mechanism (Fig.~\ref{fig:3}b, $t=0.5,1.0$).
One can see that at times $t>0.5$, the lower instability~$\#1$ can no longer 
develop independently: its range of unstable waves expands dramatically 
due to longer waves, which can be explained by the influence 
of the instability~$\#2$ developing in the same range of long waves. 
Fig.~\ref{fig:3} demonstrates also that the wavelength ratio of the fastest 
growing disturbances varies with time, both due to the diffusive expansion 
of the density pockets and their mutual influence.

{\it Nonlinear dynamics}. -- We now discuss the results 
of direct numerical simulation of the problem~(\ref{eq:1}-\ref{eq:10}). 
To see a non-linear development 
of the disturbances, the problem has been solved numerically
by a finite-difference method (for more details, see~\cite{bra16}). 
To identify the features of the quasi-periodicity of the pattern, 
we have chosen a long integration domain $H=20$, $L=125$ and 
applied the boundary conditions~(\ref{eq:8}) which exclude the imposed 
periodicity of the resulting structure. The calculations were performed at uniform 
rectangular mesh 200 by 625. As the initial condition, we use a random field 
of the stream function with amplitude less than $10^{-3}$. 

The nonlinear evolution of the system can be best understood when 
studying the changes in the terms of the density given by~(\ref{eq:6}).
Figure~\ref{fig:4} shows a consecutive restructuring of the density field over time.
At the very beginning, the base state is absolutely stable, and the fluid 
convection is absent (Fig.~\ref{fig:4}, $t=0.1$).
However, two zones of low density are already clearly visible in the figure.
The instability in the lower band is excited first. 
The pattern is a perfectly periodic system of chemoconvective cells enclosed 
between the layers of the motionless fluid (Fig.~\ref{fig:4}, $t=0.5$).
At this point in time, the wave number of the structure is about $k_1\approx 2.46$, 
which agrees well with the linear stability analysis (see Fig.~\ref{fig:3}).
Then, the instability $\#2$ is excited in the upper density pocket. 
The wavenumber of this structure is much smaller: $k_2\approx 0.5$
(Fig.~\ref{fig:4}, $t=5$). When the latter instability develops sufficiently, 
it starts to affect the chemoconvection in the lower density pocket
by injecting fresh acid with a spatial periodicity of $2\pi/k_2$.
This leads to the formation of an obvious spatial quasi-periodic pattern below 
the reaction front (Fig.~\ref{fig:4}, $t=5$ and $10$).
Finally, the pattern loses its regularity and is destroyed (Fig.~\ref{fig:4}, $t=15$).

\begin{figure}
\includegraphics [scale=0.18] {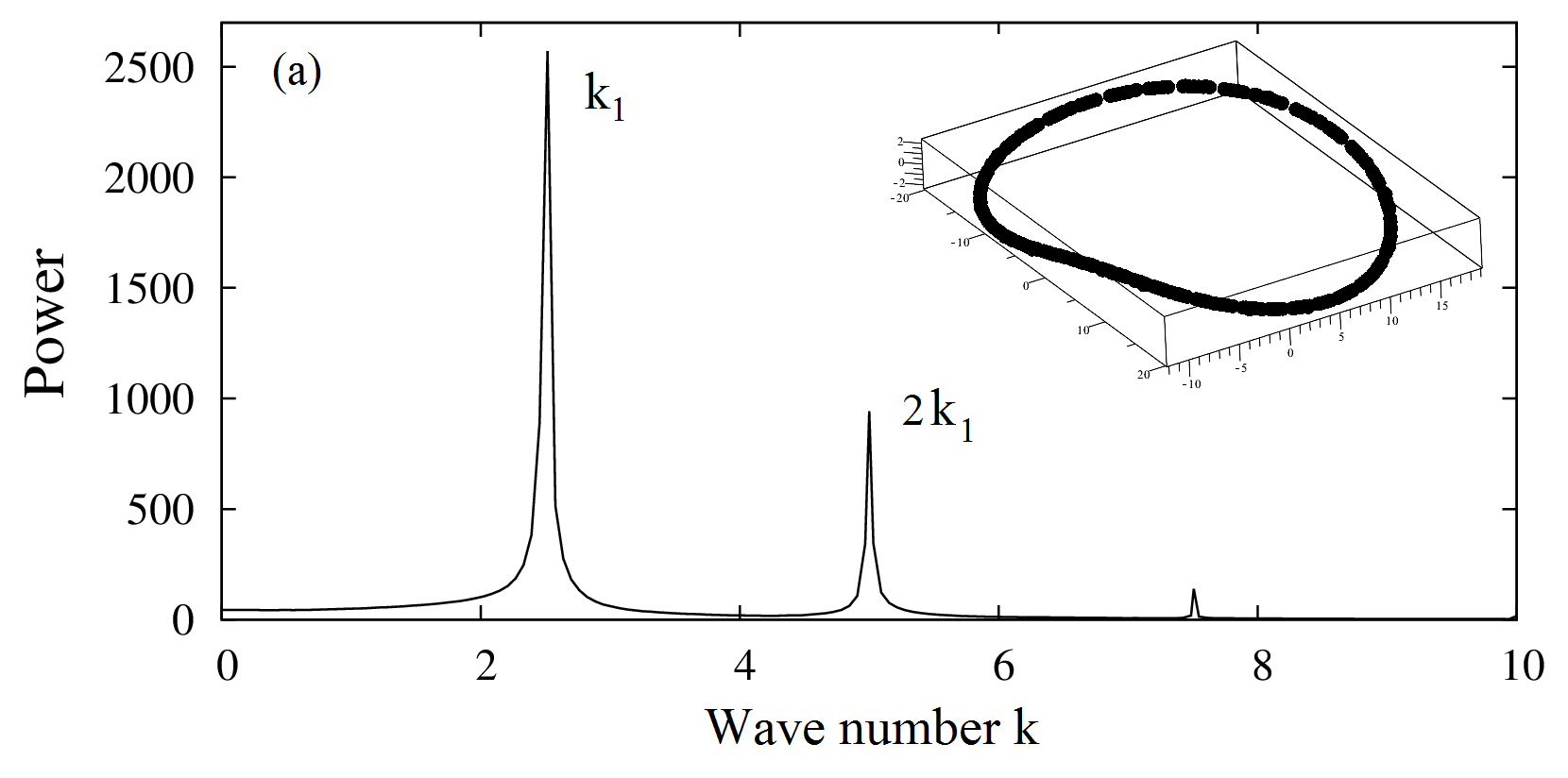}\\
\includegraphics [scale=0.18] {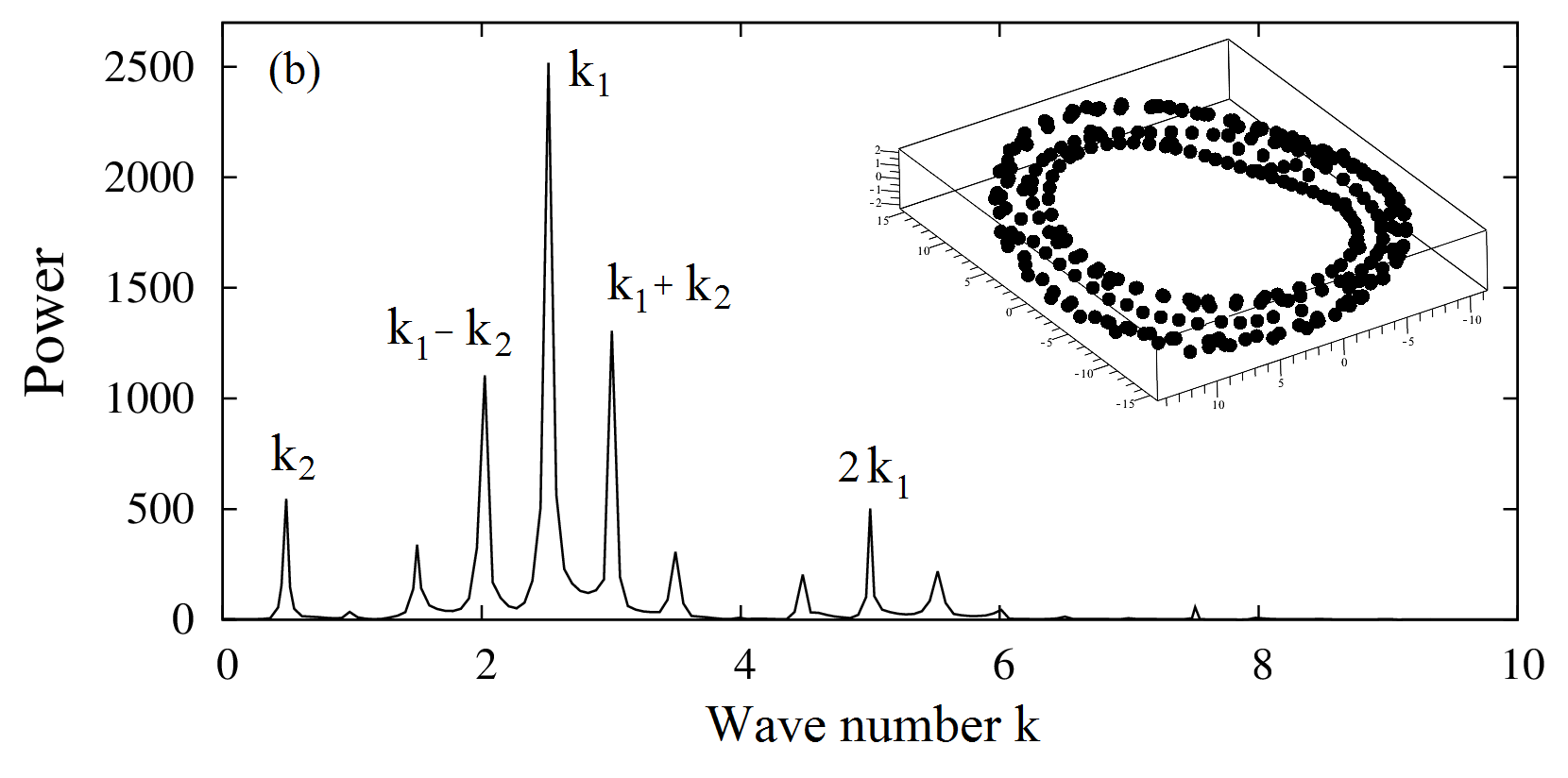}\\
\includegraphics [scale=0.18] {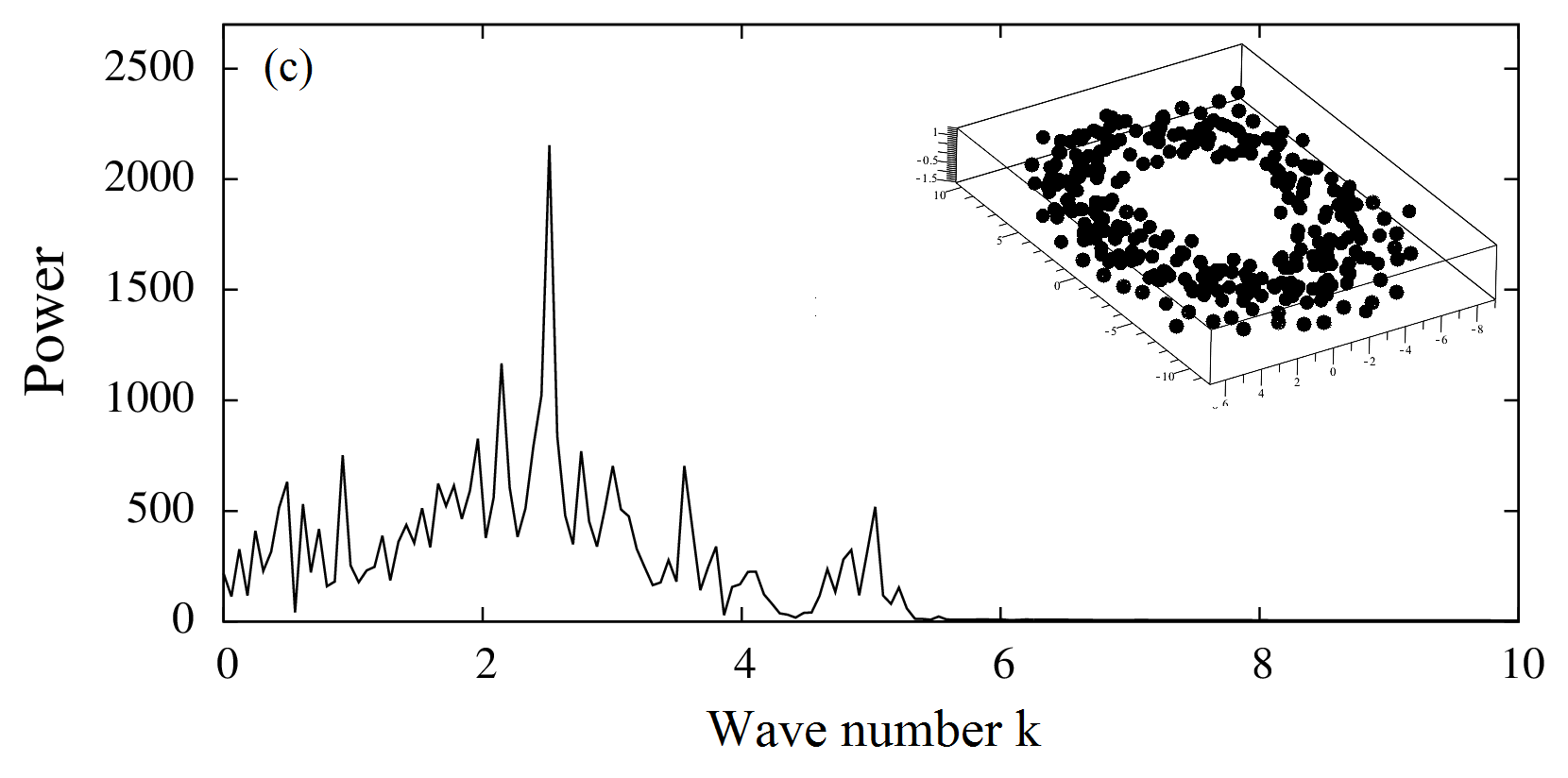}\\
\caption{\label{fig:5} 
The power spectra and phase portraits (in {\it Inset}) 
reconstructed from the averaged density ${\hat\rho}(t,x)$
at three consecutive times: (a) $t=0.5$; (b) $t=5$; (c) $t=15$.}
\end{figure}

Let us define the coordinate $x$ as a new effective time and 
consider the "dynamics" of the system. 
In addition to calculating the power spectrum, we perform 
the technique of the phase portrait reconstruction including 
the method of delays~\cite{pack80} with 
preprocessing using the singular value decomposition (SVD) 
method~\cite{broom86}. In this technique, a multi-dimensional 
embedding space is constructed from the time series data. 
The usage of the SVD method allows calculating
an optimal basis for the projection of the reconstructed phase 
dynamics of the system. In our case, this analysis can be carried out only 
for a limited number of points (625) equal to the number 
of grid nodes along $x$.
The signal for such an analysis was prepared as follows.
The density field $\rho(t,x,z)$ has been 
spatially averaged across the lower instability band 
\begin{equation}
{\hat\rho}(t,x) = \frac{1}{H_{bot}}\int_{-H_{bot}}^{0} \rho(t,x,z) dz  ,
\label{eq:16}
\end{equation}
to yield the averaged profile ${\hat\rho}(t,x)$ 
depending on time $t$ being the governing parameter 
and the longitudinal coordinate $x$ playing the role of effective time. 
Here, $H_{bot}$ stands for the width of the lower instability zone.

Fig.~\ref{fig:5} shows the Fourier spectra and phase portrait reconstructions 
calculated for three characteristic points in time. 
The dynamic mode in Fig.~\ref{fig:5}, $t=0.5$ can be unambiguously 
characterized as periodic. The first peak of the power spectrum 
determines the authentic wave number of the 1st instability $k_1\approx 2.46$, 
and the second one is simply the double value of the first peak. 
Thus, the transition from a stable base state (Fig.~\ref{fig:4}, $t=0.1$) 
to a periodic system of chemoconvective cells (Fig.~\ref{fig:4}, $t=0.5$) 
can be interpreted as a spatial Hopf bifurcation giving rise to a limit cycle
shown in {\it Inset}. The following dynamic mode demonstrates obvious signs of a quasiperiodic behavior (Fig.~\ref{fig:5}, $t=5$).
The effect of the instability~$\#2$ is expressed in the fact that the power 
spectrum now contains two characteristic peaks and all other peaks are 
just their linear combinations. The peak in the long-wave part 
of the spectrum corresponds to the wave number $k_2\approx 0.5$.
Thus, the transition from the periodic cells (Fig.~\ref{fig:4}, $t=0.5$) 
to a spatially quasi-periodic system of chemoconvective cells (Fig.~\ref{fig:4}, $t=5$) 
can be interpreted as a secondary Hopf bifurcation giving rise 
to a two-dimensional torus shown in {\it Inset} of Fig.~\ref{fig:5}, $t=5$.
Since the ``time series'' based on $\hat\rho(t,x)$ is insufficient for studying 
the complex bifurcations that require long data sequences, 
we cannot assert whether another Hopf bifurcation to a three-dimensional 
torus is occurring reliably. But we can insist that the torus eventually 
collapses giving way to a toroidal strange attractor (Fig.~\ref{fig:4}, $t=15$).
Here the spectrum contains already continuous ranges of the excited waves 
characteristic of chaotic behavior.

{\it Conclusions}. -- As it is well known, 
the Ruelle-Takens-Newhouse theory states that a typical transition to chaos 
involves two consecutive Hopf bifurcations resulting in a two-dimensional torus. 
As soon as the third Hopf bifurcation occurs, the broadband noise characteristic 
of a strange attractor should start to appear. However, this scenario 
is universal only for dynamic systems that evolve over time.
In this paper, we demonstrate that exactly the same scenario 
can also be realized if the system evolves over space.
This occurs in the physical system of two miscible solutions of HNO$_3$/ NaOH, 
in which reaction-diffusion-convection processes lead to the appearance 
of spatially separated chemoconvective instabilities interacting 
with each other by means of diffusive signals. Since the instability 
wavelengths are independent parameters, the nonlinear interaction 
of the two modes leads to the appearance of a spatial quasiperiodic pattern, 
which is destroyed in accordance with the Ruelle-Takens-Newhouse scenario.

We gratefully acknowledge the support of this work 
by the Russian Science Foundation Grant No. 19-11-00133.

\end{document}